\documentclass[prl,aps,twocolumn,showpacs]{revtex4}

\usepackage{amsmath}
\usepackage{bm}
\usepackage{graphicx}

\begin{document}

\title{Ferrofluidity in a two-component dipolar Bose-Einstein Condensate}

\author{Hiroki Saito$^1$}
\author{Yuki Kawaguchi$^2$}
\author{Masahito Ueda$^{2,3}$}
\affiliation{
$^1$Department of Applied Physics and Chemistry, University of
Electro-Communications, Tokyo 182-8585, Japan \\
$^2$Department of Physics, University of Tokyo, Tokyo 113-0033, Japan \\
$^3$ERATO Macroscopic Quantum Project, JST, Tokyo 113-8656, Japan
}

\date{\today}

\begin{abstract}
It is shown that the interface in a two-component Bose-Einstein condensate
(BEC) in which one component exhibits a dipole-dipole interaction
spontaneously forms patterns similar to those formed in a magnetic liquid
subject to a magnetic field.
A hexagonal pattern, hysteretic behavior, and soliton-like structure are
numerically demonstrated.
A phenomenon similar to the labyrinthine instability is also found.
These phenomena may be realized using a $^{52}{\rm Cr}$ BEC.
The periodic density modulation in the superfluid ground state offers
evidence of supersolidity.
\end{abstract}


\maketitle

When a magnetic liquid (a colloidal suspension of magnetic fine particles)
is subjected to a magnetic field perpendicular to the surface, the liquid
is magnetized and the surface undergoes spontaneous deformation into
characteristic patterns shaped like `horns' growing from the liquid.
This surface instability is known as the normal-field or Rosensweig
instability~\cite{Cowley} and is the subject of active
research~\cite{Rosensweig,Gailitis,Bacri,Allais,Lange,Fried,Richter}.
The system also exhibits a variety of phenomena, such as hysteretic
behaviors~\cite{Bacri}, transition between hexagonal and square
patterns~\cite{Allais}, and stabilization of a soliton-like
structure~\cite{Richter}.
Because of the visual appeal of the pattern formation, the dynamics
of the magnetic-liquid surfaces are introduced even in art~\cite{Kodama}.

\begin{figure}[t]
\includegraphics[width=8.0cm]{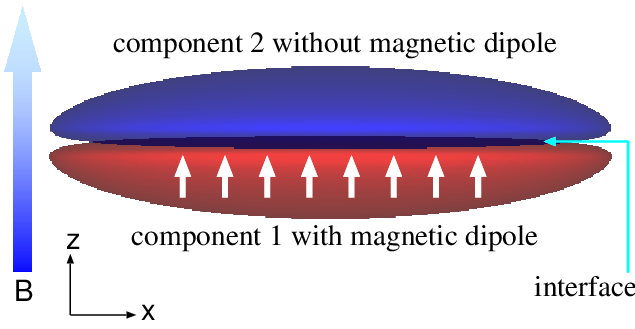}
\caption{
(Color) Schematic illustration of the system.
A two-component BEC is confined in a trapping potential, in which
component 1 has a magnetic dipole moment and component 2 does not.
The magnetic dipole is polarized in the $z$ direction by the magnetic
field.
The two components are phase-separated due to the field gradient $dB / dz
< 0$.
}
\label{f:schem}
\end{figure}

In the present Letter, we show that a Bose-Einstein condensate (BEC) of an
atomic gas with a strong dipole-dipole interaction~\cite{Gries} exhibits
instabilities and pattern formations similar to those in magnetic liquids.
The system considered here is schematically illustrated in
Fig.~\ref{f:schem}.
A two-component BEC is used, in which the atoms in component 1 have a
magnetic dipole moment and the atoms in component 2 are nonmagnetic.
A magnetic field is applied to fix the direction of the fully polarized
magnetic dipole (say, in the $z$ direction).
A magnetic-field gradient in the $z$ direction pulls the atoms in
component 1 in the $-z$ direction, and then the two components are
phase-separated for a strong field gradient.
The surface deformation in a magnetic liquid by the Rosensweig instability
occurs when it lowers the sum of gravitational, surface, and magnetic
energies~\cite{Gailitis}.
In analogy with a magnetic-liquid system, the field gradient plays the
role of gravity, and the quantum pressure and contact interatomic
interactions create an interface energy, corresponding to the surface
tension of a magnetic liquid.

A two-component BEC is used because in order for the density pattern to be
formed in a single-component BEC, the dipole-dipole interaction must
overcome the contact interaction, which leads to dipolar
collapse~\cite{Lahaye08}.
In fact, the parameter region in which a biconcave density pattern is
formed is in the immediate vicinity of the region of the collapse
instability~\cite{Goral,Wilson}.
In Refs.~\cite{Pedri,Tikho}, quasi-two dimensional (2D) systems are
considered in order to avoid the collapse, and 2D solitons are shown to be
stabilized by the dipolar interaction.
Using a multi-component BEC is another possible way of realizing pattern
formation without suffering a dipolar collapse, since texture formation is
much easier than density-pattern formation.
For example, spin domains can be formed by a small spin-dependent
interaction~\cite{Stenger}.

In this Letter we show that a hexagonal pattern of density peaks emerges
on the interface between nonmagnetic and fully polarized $^{52}{\rm Cr}$
BECs~\cite{Gries}.
We find that the number of peaks can be controlled by the strength of the
field gradient.
Hysteresis also occurs between flat and single-spike interfaces, which
resembles soliton-like structures in magnetic liquids~\cite{Richter}.
We show that various patterns are formed, including a labyrinthine
pattern~\cite{Roman}.

We consider a two-component BEC described by the macroscopic wave functions
$\psi_1$ and $\psi_2$ in the zero-temperature mean-field approximation.
The wave functions obey the nonlocal Gross-Pitaevskii (GP) equations given
by
\begin{subequations} \label{GP}
\begin{eqnarray}
i \hbar \frac{\partial \psi_1}{\partial t} & = & \biggl( -\frac{\hbar^2
\nabla^2}{2 M_1} + V_1 - \mu \frac{dB}{dz} z + g_{11} |\psi_1|^2 + g_{12}
|\psi_2|^2 
\nonumber \\
& & + \int U(\bm{r} - \bm{r}') |\psi_1(\bm{r}')|^2 d\bm{r}' \biggr)
\psi_1,
\\
i \hbar \frac{\partial \psi_2}{\partial t} & = & \left( -\frac{\hbar^2
\nabla^2}{2 M_2} + V_2 + g_{22} |\psi_2|^2 + g_{12} |\psi_1|^2 \right)
\psi_2,
\end{eqnarray}
\end{subequations}
where $M_n$ and $V_n$ are atomic mass and trap potential for component
$n = 1$ or $2$.
The interaction coefficients are given by $g_{nn'} = 2 \pi \hbar^2 a_{nn'}
/ M_{nn'}$ with $a_{nn'}$ and $M_{nn'}$ being the $s$-wave scattering
length and the reduced mass between components $n$ and $n'$. 
The dipole-dipole interaction has the form $U(\bm{r}) = \mu_0 \mu^2 (1 -
3 z^2 / r^2) / (4 \pi r^3)$, where $\mu_0$ is the permeability of vacuum
and $\mu$ is the magnetic dipole moment of the atoms in component 1.
For simplicity, we assume that the two components have the same mass $M$
and the same number of atoms $N / 2$ and that they experience the same
axisymmetric harmonic potential given by $V_n = M / 2 [\omega_\perp^2 (x^2
+ y^2) + \omega_z^2 z^2]$, where $\omega_\perp$ and $\omega_z$ are radial
and axial trap frequencies.
Gravity only shifts the origin and can be neglected.
The magnetic field depends only on $z$, and $dB / dz$ is uniform.
We employ the $m_J = -3$ magnetic sublevel of a $^{52}{\rm Cr}$ atom for
component 1, for which $\mu = 6 \mu_{\rm B}$ with $\mu_{\rm B}$ being the
Bohr magneton and $a_{11} \simeq 100 a_{\rm B}$~\cite{Werner} with $a_{\rm
B}$ being the Bohr radius.
We assume $a_{11} \simeq a_{22} \simeq a_{12}$.

We first study stable states of the system, obtained by replacing $i$ with
$-1$ on the left-hand sides of Eq.~(\ref{GP}).
The numerical propagation is performed using the Crank-Nicolson scheme and
the dipolar part is calculated using a fast Fourier transform.
The initial state of the imaginary-time propagation is a flat-interface
state $\Psi_n$ for a sufficiently large $|dB / dz|$ with small initial
noise $\psi_n = \Psi_n + r_n$, where $r_n$ represents a small complex
number randomly chosen on each mesh.
The small noise is added to break the axisymmetry of the system.

\begin{figure}[t]
\includegraphics[width=8.0cm]{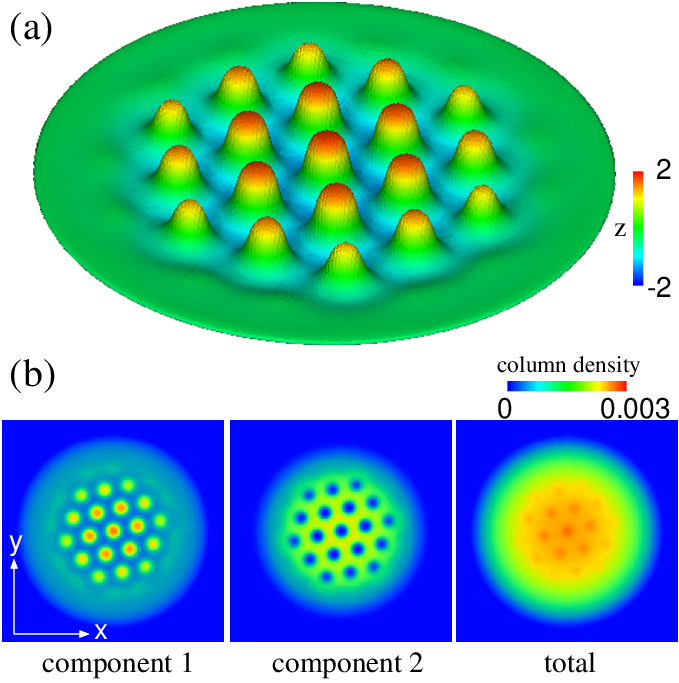}
\caption{
(Color) Stable state of the two-component dipolar BEC for $a_{11} = a_{22}
= a_{12} = 100 a_{\rm B}$ and $\mu = 6 \mu_{\rm B}$ in an axisymmetric
potential with $(\omega_\perp, \omega_z) = 2\pi \times (100, 800)$ Hz.
The number of atoms is $N = 4 \times 10^6$ with an equal population in
each component and the field gradient is $dB / dz = -600$ ${\rm mG} / {\rm
cm}$.
(a) Isodensity surface of component 1.
The color represents the $z$ coordinate normalized by $(\hbar / M
\omega_\perp)^{1/2}$.
(b) Column densities $\int |\psi_1|^2 dz$, $\int |\psi_2|^2 dz$, and $\int
(|\psi_1|^2 + |\psi_2|^2) dz$ normalized by $N M \omega_\perp / \hbar$.
The field of view is $50 \times 50$ $\mu {\rm m}$.
}
\label{f:spikes}
\end{figure}
Figure~\ref{f:spikes} (a) shows an isodensity surface of component 1
for $a_{11} = a_{22} = a_{12} = 100 a_{\rm B}$ and $dB / dz = -600$ ${\rm
mG} / {\rm cm}$.
The hexagonal mountain-like pattern on the interface between the two
components is similar to the Rosensweig pattern on a magnetic-liquid
surface.
Since the density is high and hence the dipole-dipole interaction is large
at the center of the trap, the peaks around the center are higher than
those on the periphery.
Figure~\ref{f:spikes} (b) shows the column-density profiles integrated
along the $z$ axis.
It is interesting that the left and middle panels of Fig.~\ref{f:spikes}
(b) exhibit phase separation in the $x$-$y$ plane, even though the
scattering lengths do not satisfy the immiscible condition.
This is because the density of component 1 is large at the peaks due to
the dipole-dipole interaction, which repels component 2.
In fact, the total density [right panel of Fig.~\ref{f:spikes} (b)] also
has a hexagonal pattern.

\begin{figure}[t]
\includegraphics[width=8.0cm]{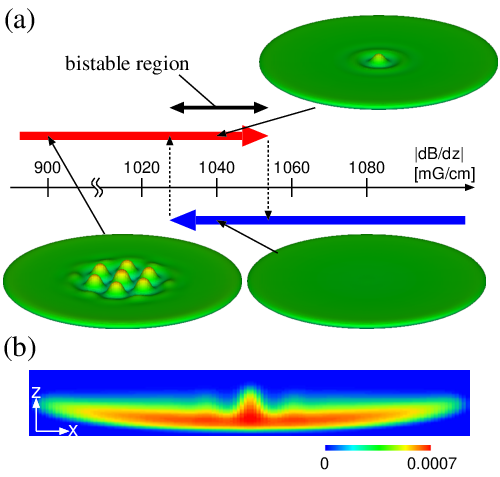}
\caption{
(Color) (a) Hysteresis with respect to a change in the field gradient.
The red and blue arrows indicate that a single-peak or flat interface is
stable for $|dB / dz| \lesssim 1054$ ${\rm mG} / {\rm cm}$ or $|dB / dz|
\gtrsim 1028$ ${\rm mG} / {\rm cm}$.
Beyond these thresholds, the two stable states convert to each other
(dotted arrows).
The isodensity surfaces are shown for $dB / dz \simeq -900$ and
$-1040$ ${\rm mG} / {\rm cm}$.
(b) Density profile of the cross section on the $y = 0$ plane for the
single-peak state at $dB / dz \simeq -1040$ ${\rm mG} / {\rm cm}$.
The field of view is $44 \times 6.7$ $\mu {\rm m}$ and the density is
normalized by $N (M \omega_\perp / \hbar)^{3/2}$.
The parameters are the same as those in Fig.~\ref{f:spikes} except for $dB
/ dz$.
}
\label{f:hysteresis}
\end{figure}

When the field gradient becomes tight, the number and height of the peaks
decrease and eventually the interface becomes flat.
The appearance and disappearance of the pattern exhibits hysteresis with
respect to a change in the field gradient.
Figure~\ref{f:hysteresis} (a) depicts stable states obtained from the
imaginary-time propagation of the GP equation as $|dB / dz|$ is increased
or decreased.
When $|dB / dz|$ is increased from a small value [red arrow in
Fig.~\ref{f:hysteresis} (a)], the hexagonal pattern of the peaks changes
into a single peak, which then disappears at $dB / dz \simeq -1054$
${\rm mG} / {\rm cm}$.
On the other hand, when $|dB / dz|$ is decreased from a large value [blue
arrow in Fig.~\ref{f:hysteresis} (a)], the interface remains flat until
$dB / dz \simeq -1028$ ${\rm mG} / {\rm cm}$.
Therefore, there is a bistable region at $|dB / dz| \simeq 1028$-$1054$
${\rm mG} / {\rm cm}$, in which the single-peak state and the
flat-interface state are both stable.
In the bistability region, the energies of the two states are almost
degenerate.
Such hysteretic behavior has been predicted~\cite{Gailitis,Fried}
and observed~\cite{Bacri,Richter} for magnetic liquids, in which the
applied magnetic field and hence the magnetization is changed.
For a wider pancake-shaped BEC, bistability between the flat-surface and
hexagonal-peak states may be observed as in a magnetic liquid.

The single-peak structure in the bistable region in
Fig.~\ref{f:hysteresis} (a) is similar to the
``ferrosoliton''~\cite{Richter} in a magnetic liquid, which shows a
stable soliton-like peak.
Figure~~\ref{f:hysteresis} (b) shows the density profile of the cross
section of the single-peak state at $dB / dz \simeq -1040$ ${\rm mG} /
{\rm cm}$.
Around the peak, the interface oscillates in a concentric manner, as in
the ferrosoliton in a magnetic liquid~\cite{Richter}.

\begin{figure}[t]
\includegraphics[width=8.0cm]{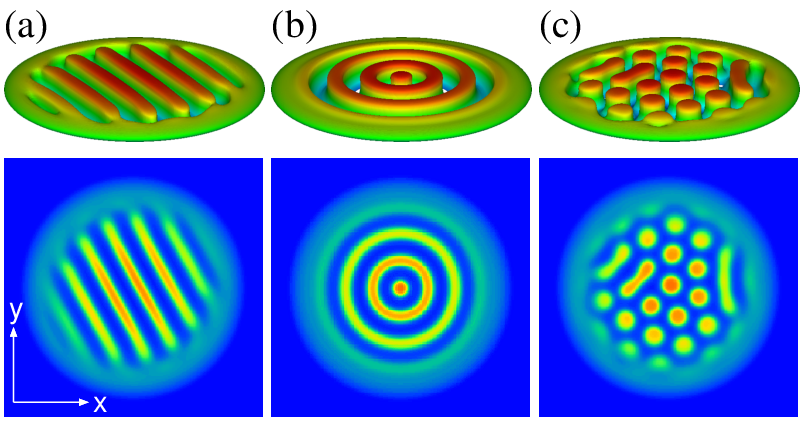}
\caption{
(Color) Isodensity surface (upper figures) and column density (lower
panels) of component 1 for various stable states.
The difference in (a)-(c) is only in the initial perturbations in the
imaginary-time propagation.
The field gradient is $dB / dz = -200$ ${\rm mG} / {\rm cm}$.
The other parameters and the color scales are the same as those in
Fig.~\ref{f:spikes}.
}
\label{f:various}
\end{figure}
To investigate various metastable patterns, we apply initial perturbations
to trigger the pattern formation.
Figure~\ref{f:various} (a) is obtained by applying an additional magnetic
field $\propto \sin(k_x x + k_y y)$ for $t < 1$ ms in the imaginary-time
propagation, where $k_x a_\perp = \sqrt{3}$ and $k_y a_\perp = 1$ with
$a_\perp = (\hbar / M \omega_\perp)^{1/2}$.
We find that the imaginary-time propagation relaxes the system to the
stripe pattern as shown in Fig.~\ref{f:various} (a).
This pattern is robust against small additional noise.
The initial temporary magnetic field for Fig.~\ref{f:various} (b) is
$\propto \cos (\sqrt{x^2 + y^2} / a_\perp)$.
The concentric pattern in Fig.~\ref{f:various} (b) is stable against
axisymmetry-breaking noise.
Figure~\ref{f:various} (c) is obtained without an additional magnetic
field, with only a small noise added to the initial state.
These results imply that there are many metastable states with various
patterns.
In Fig.~\ref{f:various} (c), the peaks partly merge into each other.
The merging of the peaks tends to occur for a large dipole interaction
and small field gradient.
For $dB / dz = -600$ ${\rm mG} / {\rm cm}$, the stripe and concentric
patterns in Figs.~\ref{f:various} (a) and \ref{f:various} (b) are unstable
against forming peaks.
Although the square lattice of the peaks is found to be unstable for the
parameters in Figs.~\ref{f:spikes} and \ref{f:various},
a hexagonal-square-lattice transition may occur for other parameters, as
in magnetic liquids~\cite{Gailitis,Allais}.

\begin{figure}[t]
\includegraphics[width=8.0cm]{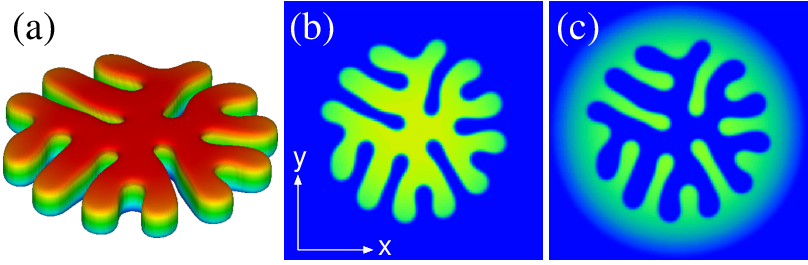}
\caption{
(Color) (a) Isodensity surface of component 1 and column density of
(b) component 1 and (c) component 2 for $a_{11} = 100 a_{\rm B}$, $a_{12}
= a_{22} = 120 a_{\rm B}$, $dB / dz = 0$, and $N = 8 \times 10^6$.
The other conditions and the color scales are the same as those in
Fig.~\ref{f:spikes}.
}
\label{f:labyrinth}
\end{figure}
Figure~\ref{f:labyrinth} shows a stable state for $dB / dz = 0$ which has
an intricate pattern similar to the labyrinthine pattern in a thin layer
of magnetic liquid confined with an immiscible nonmagnetic
liquid~\cite{Roman}.
The relation between the labyrinthine instability in a magnetic
liquid~\cite{Rosensweig} and that in the present system remains to be
clarified.

\begin{figure}[t]
\includegraphics[width=8.0cm]{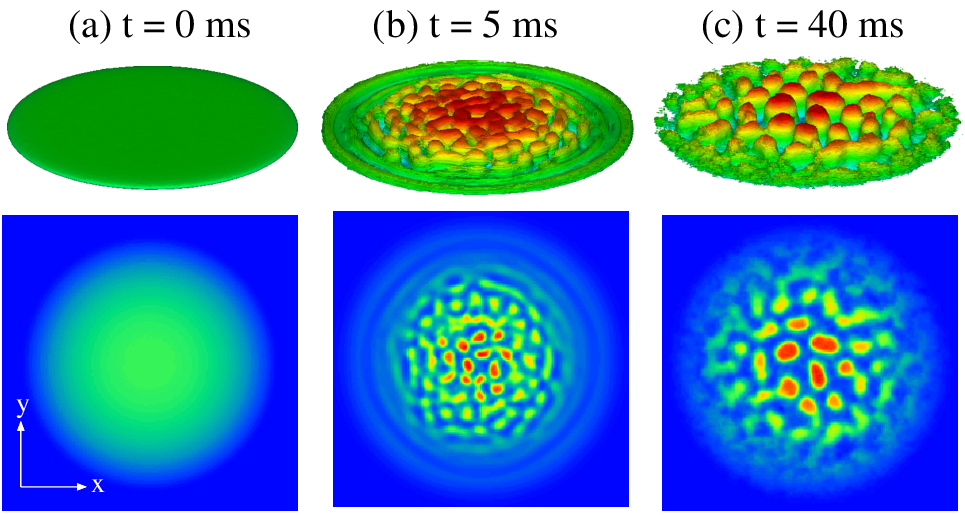}
\caption{
(Color) Time evolution of the isodensity surface (upper figures) and
column density (lower panels) of component 1.
The initial state is the flat-interface state for $dB / dz = -1200$ ${\rm
mG} / {\rm cm}$ and the field gradient changes to $dB / dz = -600$
${\rm mG} / {\rm cm}$ at $t = 0$.
The field of view is $50 \times 50$ $\mu {\rm m}$.
The parameters and color scales are the same as those in
Fig.~\ref{f:spikes}.
}
\label{f:realtime}
\end{figure}

Next we study the dynamics of the pattern formation in a nondissipative
system by solving the real-time propagation of Eq.~(\ref{GP}).
Figure~\ref{f:realtime} shows the time evolution of the system, where the
initial state is a flat-interface state for $dB / dz = -1200$ ${\rm mG}
/ {\rm cm}$, as shown in Fig.~\ref{f:realtime} (a).
In order to break the axisymmetry, a small initial noise is added to the
initial state.
We let the system evolve in time with $dB / dz = -600$ ${\rm mG} / {\rm
cm}$, for which the stable state has hexagonal peaks, as shown in
Fig.~\ref{f:spikes} (a).
The density fluctuations grow into many peaks, as in Figs.~\ref{f:spikes}
(b) and \ref{f:spikes} (c), due to the Rosensweig instability.
We note that the characteristic wavelength in the pattern at the early
stage [Fig.~\ref{f:spikes} (b)] is smaller than that in the later pattern
[Fig.~\ref{f:spikes} (c)].
This difference indicates that the most unstable wavelength in the linear
regime does not correspond to the final stable pattern determined by the
nonlinear interaction.
A similar situation also occurs in magnetic liquids~\cite{Lange}.
If the phenomenological dissipation is taken into account by replacing $i$
with $i - 0.03$~\cite{Tsubota} in Eq.~(\ref{GP}), the hexagonal pattern is
formed at $\sim 50$ ms.

A candidate for component 2 is the $^7{\rm S}_3$ $m_J = 0$ state of a
$^{52}{\rm Cr}$ atom.
Though the $m_J \neq -3$ states are unstable against dipolar collisions,
the lifetime is long enough ($\sim$ s)~\cite{Gries} to observe the
pattern-formation dynamics.
The spin exchange dynamics, such as $m_J = 0 + 0 \rightarrow -3 + 3$, can
be suppressed using, e.g., the microwave-induced quadratic Zeeman
effect~\cite{Leslie}.
The scattering lengths measured in Ref.~\cite{Werner} give $a_{22} \simeq
60.7 a_{\rm B} + a_0 / 7$ and $a_{12} \simeq 67.8 a_{\rm B}$, with $a_0$
being the scattering length for the colliding channel with total spin
0, which is unknown.
The scattering length $a_{11}$ can be controlled using the magnetic
Feshbach resonance~\cite{Lahaye07}.
The inhomogeneity of the magnetic field $\sim 1 {\rm G} / {\rm cm} \times
10$ $\mu{\rm m}$ is much smaller than the resonance width 1.7
G~\cite{Werner}.
We can therefore obtain $a_{11} \sim a_{22} \sim a_{12} \sim 60 a_{\rm B}$
by changing $a_{11}$, if $|a_0|$ is not very large.
We have confirmed that the Rosensweig pattern emerges for these scattering
lengths.
Another possibility for component 2 is an alkali atom, for which, however,
$a_{12}$ has not been measured yet.
The pattern formation for various scattering lengths merits further
study.

An important difference between the present system and magnetic liquids
is that the present system is a superfluid, while a conventional
magnetic liquid is a normal fluid.
In the Rosensweig pattern, the total density shows a periodic modulation
(i.e., diagonal order), while the present system by construction possesses
an off-diagonal order, presenting strong evidence of supersolidity.
Another difference is that the present system is fully polarized
irrespective of the strength of the magnetic field, while the magnetic
liquid is paramagnetic.
The difference between a compressible gas and an incompressible liquid is
also important.
It is remarkable that both systems exhibit similar phenomena despite these
differences.

In conclusion, we have studied pattern formation on the interface in a
two-component BEC, in which one component exhibits a dipole-dipole
interaction.
We found a rich variety of interfacial patterns, including Rosensweig
hexagonal peaks and a labyrinthine pattern.
We also observed hysteretic behavior and the ferrosoliton, as in magnetic
liquids.
We expect that the two-component system proposed here will provide a new
insight into surface and interface physics with long-range interactions.

This work was supported by MEXT Japan
(KAKENHI No.\ 17071005 and No.\ 20540388, and the Global COE Program ``the
Physical Sciences Frontier'') and by the Matsuo Foundation.

\end{document}